\documentclass[aps,preprint,groupedaddress]{revtex4}
\usepackage{amsmath, amsthm, amssymb}

\begin{document}

\title{Jet Formation in the magnetospheres of supermassive black holes: analytic solutions describing
energy loss through Blandford-Znajek processes}
\author{Govind Menon}
\address{Dept.\ of Mathematics and Physics,\\
Troy University, Troy, Alabama, 36082}
\author{Charles D.\ Dermer}
\address{E. O. Hulburt Center for Space Research, Code 7653,
Naval Research Laboratory, Washington, DC 20375-5352}
\date{\today}
\begin{abstract}
In this paper, we provide exact solutions for the extraction of energy from a rotating black hole via both the electromagnetic Poynting flux and matter currents. By appropriate choice of a
radially independent poloidal function $\Lambda(\theta)$, we find solutions where the dominant outward energy flux is along the polar axis, consistent with a jet-like collimated outflow, but also
with a weaker flux of energy along the equatorial plane.  Unlike all the previously obtained solutions (\cite{Ref1}, \cite{Ref8}, \cite{Ref11}), the magnetosphere is free of magnetic monopoles everywhere.
\end{abstract}


\maketitle Jet Formation And Energy Extraction

\section{Introduction}

In 1977, \cite{Ref1} introduced the force-free, stationary, axisymmetric magnetosphere of the Kerr geometry as a possible setting for the extraction of energy from supermassive black holes at an astronomical scale. To this day, astrophysicists consider the Blandford-Znajek process as the leading mechanism for the observed phenomenon of luminous black holes \cite[for example,][]{Ref10,Ref3,Ref9}. Indeed, it is this mechanism we have focused on as well \citep{Ref8,Ref11,MD09,DMbook}.

In an earlier paper \citep{Ref11}, we were sucessful in providing the only known class of exact analytic solutions to the equations of force-free electrodynamics in Kerr geometry. Although finite everywhere in the magnetosphere, these solutions did not appear to be physically realistic. In particular, the currents in the magnetosphere were null vectors fields suggesting that the charged particles traveled at the speed of light. Additionally, the current vector field was inward pointing resulting in an influx of electromagnetic energy. This seemed to suggest that the solutions were not physically interesting.

In this paper, we provide a clear, physically realistic interpretation of the
current density vector field. In particular, we decompose the infalling null vector field to  currents each of which have future pointing timelike velocities as these are candidates for physically realistic currents. In this decomposition, one of the currents is outgoing. This current provides a concrete mechanism for jet formation in black holes. However, the electromagnetic flux continues to be inward pointing. To remedy this, we extend the results via a symmetry transformation. Briefly, it turned out that the net current vector is proportional to the infalling  principle null geodesic of the Kerr geometry. Since the Kerr geometry is of Petrov type D \citep{CH83}, it is only natural to ask whether the equations of electrodynamics would allow the existence of solutions where the current vector was proportional to  the second principle (outgoing) null geodesic of the Kerr geometry. As we shall show in this paper, this is indeed the case. And the existence of this dual class of solutions is a general property of the equations, not necessarily dependent on our particular solution. The dual solution to our particular solution (hereby referred to as the $\Omega_-$ solution) does allow for the extraction of energy via  the electromagnetic Poynting flux.

All generalities are restricted by picking a concrete example and carrying out the energy extraction rates from matter currents and the electromagnetic Poynting flux. Here we find that the matter current  and the electromagnetic Poynting flux naturally describe a polar jet for a specific choice of an arbitrary poloidal function $\Lambda$. This solution
also describes a secondary local maxima in the energy extraction rate peaking near the equatorial plane, which might correspond to outflowing Poynting flux that could drive a disk wind.
Even though our past $\Omega_+$ solution \citep{Ref8} generalized the Blandford-Znajek split monopole solution, the non-existence of a magnetic monopole for our new solution is shown here  by direct computation \cite[this was not the case in the original approximate solution presented by] []{Ref1}.

\section{The $\Omega_-$ Solution}

In \cite{Ref11} we derived the following class of exact solutions for the force-free magnetosphere of the Kerr black hole. Here, the components of the electromagnetic fields in the Boyer-Lindquist coordinate system are given by

\begin{equation}
E_\varphi=0=E_r \;,
\end{equation}
\begin{equation}
E_\theta =-\frac{2}{a^2}\; \Lambda \;\frac{\cos{\theta}}{\sin^5{\theta}}\;,
\end{equation}
and
\begin{equation}
B^\theta=0\;,
\end{equation}
\begin{equation}
B^r=\alpha\;H^r = \frac{2}{a} \;\Lambda \;\frac{\cos{\theta}}{\sqrt{\gamma}\;\sin^3{\theta}}\;,
\label{radb}
\end{equation}
and
\begin{equation}
\alpha\;B_\varphi = H_\varphi=\frac{2}{a^2}\; \Lambda\; \frac{\cos\theta}{\sin^4\theta}\;.
\label{finalhphi}
\end{equation}
Finally,
\begin{equation}
\Omega_- = \frac{1}{a \sin^2 \theta}
\end{equation}
Please see Appendices \ref{KGE} and  \ref{EESS} for the definitions of the quantities listed above.
The general subscripts $1,2,3$ in the Maxwell tensor corresponds to $r, \theta, \varphi$ respectively in our case. Here $\Lambda$ is an arbitrary function of $\theta$.

It is only natural to expect the net current vector field to follow a geodesic under force-free conditions. This is indeed the case:
\begin{equation}
I =-\frac{2}{a^2\; \alpha\; \sqrt\gamma} \;\;\frac{d}{d \theta}\left[\Lambda \frac{\cos{\theta}}{\sin^4{\theta}}\right ]\;n\;.
\end{equation}
Here $n$ is the infalling principle null geodesic of the Kerr geometry. Explicitly,
\begin{equation}
n= \frac{r^2+a^2}{\Delta}\;\partial_t\; - \;\partial_r \;+\;\frac{a}{\Delta}\;\partial_\varphi.
\end{equation}
A simple calculation will reveal that the solution presented above satisfies Maxwell's equations (eq.(\ref{maxeq})) and the force-free condition, eq.(\ref{divtemforce}). The above solution is well defined everywhere in the magnetosphere. In particular, since our solution satisfies the Znajek regularity condition  (eq.(\ref{znaregcond})), the fields are well defined at the event horizon, as we will explicitly verify by going into the Kerr-Schild coordinate system (see subsection \ref{KSFields}). Also, the apparent singularity at the poles is removed by the transformation  $\Lambda \rightarrow \sin^5 \theta\;\Lambda$.

However, the above solution, as it stands, lacks any meaningful physical interpretation. Physically realistic charges cannot flow along null geodesics. This problem will be remedied in the remainder of this paper by decomposing the null current vector into timelike vector fields that are possible worldlines of charged particles in the magnetosphere. In  \cite{Ref11}, it was the deduction of a viable expression for $\Omega$ that immediately gave us the expressions for all the fields and currents. Therefore, we shall refer to the complete solution listed above as the $\Omega_-$ solution.
\subsection{The $\Omega_-$ Solution In The Kerr-Schild Coordinate System}
\label{KSFields}
Transforming the Maxwell tensor $F_{\mu \nu}$ into the Kerr-Schild coordinate system (see section \ref{KGE}), we see that
\begin{equation}
F_{\bar t \bar r} = F_{\bar t \bar \varphi}=  F_{\bar r \bar \varphi}=F_{\bar r \bar \theta}=0 \;,
\end{equation}
\begin{equation}
F_{\bar t \bar \theta} = - E_\theta\;,
\end{equation}
\begin{equation}
F_{\bar \theta \bar \varphi} = \sqrt{\gamma}\;B^r\;,
\end{equation}
and
$$I =\frac{2}{a^2\; \alpha \;\sqrt\gamma} \;\;\frac{d}{d \theta}\left[\Lambda \frac{\cos{\theta}}{\sin^4{\theta}}\right ]\;\partial_{\bar r}\;.$$
Thus we see that the fields and currents are well defined on the event horizon $r = r_+$ as well. This is necessarily so since we had insisted on the Znajek regularity condition given by eq.(\ref{znaregcond}) in the derivation of our solution \citep{Ref11}.
\label{PolarFields}

\subsection{The Electromagnetic Poynting Flux}
\cite{Ref1} computed the expression for the energy extracted from a rotating Kerr black hole via the electromagnetic Poynting vector for the force-free, stationary, axis-symmetric magnetosphere. In the 3+1 notation, the rate of electromagnetic energy extraction becomes
\begin{equation}
\frac{d  {\cal E}_{EM}}{dt}=-\int  H_\varphi  \Omega  B^r \sqrt {\gamma_{rr}} \;dA\;.
\label{engextractformula}
\end{equation}
For the case of our $\Omega_-$ solution, the above expressions gives that
\begin{equation}
\frac{d {\cal E}_{EM}}{dt}=-\frac{8\pi}{a^4} \int_{0}^{\pi} \frac{\Lambda^2\; \cos^2\theta}{\sin^9\theta} \;d\theta \leq 0 \;.
\label{omegaminusenergy}
\end{equation}
I.e., if stationary electromagnetic fields can indeed transfer energy to and from a black hole, the $\Omega_-$ solution in particular does not allow for energy extraction. Instead, the black hole behaves as an energy sink. We will come back to this point when we consider a dual solution.

\section{A Timelike Decomposition Of The Null Current}
\subsection{Region I}
Region I of the Kerr black hole is defined by the condition $r > r_+$, and the region of spacetime given by $r_- < r \leq r_+$ will be referred to as region II. Define vector fields $V_I$  and $W_I$ in region I by

\begin{equation}V_I = \frac{(r^2+a^2)\;\partial_t + a\; \partial_\varphi}{\sqrt{\rho^2\;\Delta}}\label{reigionIV}\end{equation}
and
\begin{equation}
W_I = \frac{ \; (3l-1)\left[(r^2+a^2)\;\partial_t + \;a\; \partial_\varphi\;\right] + \Delta \;\partial_r }{\sqrt{\rho^2\;\Delta}\sqrt{3l(3l-2)}}\;.\label{reigionIW}\end{equation}
\vskip0.2in \noindent
Here, $l \equiv l(\theta)$ is the {\it energy collimation factor} such that $l(\theta) > 2/3$ everywhere. Then,

$$g (V_I, V_I) = -1\;\;\;{\rm
and }\;\;\;\lim_{r\rightarrow \infty} g(V_I,\partial_t) = -1\;,$$
and
$$g (W_I, W_I) = -1$$
and$$\lim_{r\rightarrow \infty} g(W_I,\partial_t) = -\frac{(3l-1)}{\sqrt{3l(3l-2)}}<0\;.$$

\vskip0.2in\noindent
That is, $V_I$ and $W_I$ are future pointing timelike in region I, and are candidate proper velocities of charged particles. Therefore, in region I we can write the current vector as the flow of two oppositely charged timelike currents
$$I_I = I_{Ia} + I_{Ib} $$
where
$$I_{Ia} = -\frac{6\;l}{a^2\;\sin\theta \sqrt{\rho^2 \Delta}} \;\;\frac{d}{d \theta}\left[\Lambda \frac{\cos{\theta}}{\sin^4{\theta}}\right] \;V_I$$
and
$$I_{Ib} = \frac{2\sqrt{3l(3l-2)}}{a^2\;\sin\theta \sqrt{\rho^2 \Delta}} \;\;\frac{d}{d \theta}\left[\Lambda \frac{\cos{\theta}}{\sin^4{\theta}}\right]\;W_I \;.$$
Naturally, $I_{Ib}$ describes an outgoing current. For completeness note that in Kerr-Schild coordinates

$$V_I = \frac{\left[\; (r^2+a^2)\;\partial_{\bar t} + a\; \partial_{\bar \varphi}\; \right]}{\sqrt{\rho^2\;\Delta}}$$ and $$W_I = \frac{\left[ \; 3\;l\;(r^2+a^2)\;\partial_{\bar t} + 3\;a\;l\; \partial_{\bar \varphi}\; + \Delta \;\partial_{\bar r} \right]}{\sqrt{3\rho^2\;\Delta\;l(3l-2)}}\;.$$
\vskip0.2in\noindent
It is important to note, that in our decomposition, even in the Kerr-Schild coordinate system, $V_I$ and $W_I$ are not defined well defined at $r_+$. There is a deeper reason why this happens regardless of the decomposition if we require an outgoing current in region I. Consider an arbitrary decomposition of the net current vector into terms proportional to future pointing timelike vectors in region I. At least one of these vectors, say $T$ must be of the form:
$$T = X + \psi^2 \; \partial_{\bar r}\;,$$
where $\psi$ is an arbitrary function of spacetime coordinates and $X$ has no other radial components so that the net radial component is positive. The only future pointing causal vector  at $r_+$ such that its radial component is greater than or equal to zero is of the form
$$V_+ = d^2\;\left[(r_+ ^2 + a^2)\;\partial_{\bar t} + a \;\partial_{\bar \varphi}\right]$$
where $d \in R $ is  a constant \cite[see][]{Oneill}.
Then we have that
$$\lim_{r \rightarrow r_+} \psi^2  =0$$
and
$$d^2\;\left[(r_+ ^2 + a^2)\;\partial_{\bar t} + a \;\partial_{\bar \varphi}\right]= \lim_{r \rightarrow r_+} T = \lim_{r \rightarrow r_+} X\;, $$
for some non zero constant $d$ (since a unit timelike vector field $T$ cannot abruptly become the $0$ vector).
I.e., $T$ is null  at the event horizon (since $V_+$ is null), and hence cannot be normalized. This is the reason why $V_I$ and $W_I$ are not defined well defined at $r_+$.
\subsection{Region II }
Here, since we are including $H_+$, our decomposition must be valid at the horizon as well. Consequently, all expressions will be given in the Kerr-Schild coordinate system. Additionally, the radial component of all the vectors in this decomposition must be inward pointing to agree with the causality conditions of the interior geometry (unlike in region I where we wanted an outflowing current). As we shall see, by construction, the decomposition here will be well defined at the horizon. In region II define vector fields $V_{II}$ and $W_{II}$ by
$$V_{II} = \frac{\left[\; (r^2+a^2)\;\partial_{\bar t} + a\; \partial_{\bar \varphi} + (\Delta -2)\;\partial_{\bar r}\; \right]}{\sqrt{\rho^2\;(4-\Delta)}}$$ and $$W_{II} = \frac{\left[\; (r^2+a^2)\;\partial_{\bar t} + a\; \partial_{\bar \varphi} + (\Delta -1)\;\partial_{\bar r}\; \right]}{\sqrt{\rho^2\;(2-\Delta)}}\;.$$
Then,
$$g (V_{II}, V_{II}) = -1=g (W_{II}, W_{II})$$
and
$$\sqrt{\rho^2 (4 - \Delta)}\;\;
g(V_{II},-\partial_{\bar r}) = -\rho^2 $$$$= \sqrt{\rho^2 (2 - \Delta)}\;\;g(W_{II},-\partial_{\bar r})\;.$$
\vskip0.2in\noindent
I.e., $V_{II}$ and $W_{II}$ are future pointing timelike in region II, and are candidate proper velocities of charged particles. Therefore, in region II we can write the current vector as the flow of two oppositely charged, infalling timelike currents

$$I_{II} = I_{II\;a} + I_{II\;b} $$
where

$$I_{II\;a} = -\frac{2\sqrt{\rho^2(4- \Delta)}}{a^2\;\rho^2 \;\sin\theta } \;\;\frac{d}{d \theta}\left[\Lambda \frac{\cos{\theta}}{\sin^4{\theta}}\right] \;V_{II}$$
and

$$I_{II\;b} = \frac{2\sqrt{\rho^2(2- \Delta)}}{a^2\;\rho^2 \;\sin\theta } \;\;\frac{d}{d \theta}\left[\Lambda \frac{\cos{\theta}}{\sin^4{\theta}}\right] \;W_{II} \;.$$
\subsection{Separating The Regions Outside The Event Horizon}
\label{outeventreg}
Collectively, $I_I = I$ and $I_{II} = I $ give a meaningful description of the net current everywhere in the magnetosphere.
The regions we chose might lead us to believe that the difference in the character of our decomposition must occur at the event horizon. This is not necessarily the case. $V_{II}$ and $W_{II}$ continue to be future pointing timelike in some open interval outside  the event horizon. I.e., there exists $\delta >0$ such that $V_{II}$ and $W_{II}$  are future pointing timelike in $r \leq r_+ + \delta$ (for a fixed $\bar t$, the surface given by $\bar r = r_+$ is compact). Therefore in the region given by $r \leq r_+ + \delta$, we set
$$I = I_{II} = I_{II\;a} + I_{II\;b} $$
and for $r > r_+  +\;\delta$ we set
$$I = I_I = I_{I\;a} + I_{I\;b} \;.$$
In fact, the new regions has further advantages, in that
$$\lim_{r \rightarrow r_+ \;+ \;\delta} \;I_{I\;a}\;\;\;{\rm and}\;\;\;\lim_{r \rightarrow r_+ \;+\; \delta} \;I_{I\;b} $$ are well defined.
\section{Extraction Of Energy From Matter Currents}
Here we focus on the outgoing current in region I:

$$I_{Ib} = \frac{2\sqrt{3l(3l-2)}}{a^2\;\sin\theta \sqrt{\rho^2 \Delta}} \;\;\frac{d}{d \theta}\left[\Lambda \frac{\cos{\theta}}{\sin^4{\theta}}\right]\;W_I \;.$$
\vskip0.2in\noindent
It is easy to see that $I_{Ib}$ is divergence free.
We now use this conserved current, that is the only outflowing current in this decomposition, to construct an expression for the extraction of matter energy from the black hole. Define the charge density $\rho_c$ by
$$\rho_c = \alpha \;I_{Ib}^t$$
and the current 3-vector $\vec J$ by
$$J^i = \alpha\; I_{Ib}^i$$
for $i=1,2,3$. $\nabla_\mu \;I_{Ib}^\mu = 0$ implies that
$$\partial_t \rho_c + \tilde \nabla \cdot J = 0\;.$$
Now, assuming that at every point $I_{Ib}$ is comprised of only one species of charged particle with charge $q$ and mass $m$, we can define the mass density of the current as
$$\rho_m = \frac{m}{q}\;\rho_c $$ so that $\rho_m \geq 0$. This happens when
\begin{equation}\frac{1}{q}\;\frac{d}{d \theta}\left[\Lambda \frac{\cos{\theta}}{\sin^4{\theta}}\right] \geq 0\;.
\label{massdensitycond}\end{equation}
Then
$$\frac{d}{dt}\; M = - \left[\int_{r\rightarrow \infty}\frac{m}{q}\; g(J \cdot  n) \;dA -\int_{r\rightarrow r_+} \frac{m}{q}\; g(J \cdot  n)\;dA \right]\;.$$
\vskip0.2in\noindent
In the above equation, $n$ is the outward pointing normal, meaning  $n = \partial_r/\sqrt{g_{rr}}$ when $r\rightarrow \infty$ and $n = -\partial_r/\sqrt{g_{rr}}$ when $r\rightarrow r_+$, and $dA = \sqrt{\gamma_{\theta\theta}\gamma_{\varphi \varphi}}\; d\theta\; d\varphi$. Therefore, the matter extraction of energy ${\cal E}_M $ from the black hole is given by

\begin{equation}
\frac{d}{dt}\; {\cal E}_M =  \int_\infty \;\frac{m}{q}\;I^r _{Ib} \; \rho^2 \sin\theta \; d\theta\; d\varphi\;.
\label{matengexteqn}
\end{equation}
\vskip0.2in\noindent
For our particular current $I _{Ib}$, we get

$$\frac{d}{dt}\; {\cal E}_M= \frac{4\;\pi}{a^2}\;\int \; \frac{m}{q}\;\frac{d}{d \theta}\left[\Lambda \frac{\cos{\theta}}{\sin^4{\theta}}\right]\; \; d\theta$$
\vskip0.2in\noindent
which from eq.(\ref{massdensitycond}) is greater than zero, thus allowing for extraction of matter energy from the black hole.
Of course, the above equation is meaningful only when $\Lambda$ is able to absorb the infinity produced by $(
\sin^4 \theta)^{-1}$, which is an easy task since $\Lambda$ is any arbitrary function of $\theta$. To understand the collimation effects, it is important to note that

$$\frac{d^2{\cal E}_M}{dt\;dA} =  \frac{m}{q}\; \vec J \cdot \vec n =  \frac{m}{q}\;\frac{2}{a^2}\;\frac{d}{d \theta}\left[\Lambda \frac{\cos{\theta}}{\sin^4{\theta}}\right] \;\frac{1}{\sqrt{\Sigma^2}\sin\theta}$$
$$ \approx  \frac{m}{q}\;\frac{2}{a^2}\;\frac{d}{d \theta}\left[\Lambda \frac{\cos{\theta}}{\sin^4{\theta}}\right] \;\frac{1}{r^2\sin\theta}\;.$$
\vskip0.2in\noindent
Therefore, we see that $(\sin\theta)^{-1}$ factor gives the currents a preferential polar jet like feature. Here, unlike the case of the stationary electromagnetic fields, the mechanism by which charged particles carry energy from the black hole is apparent from the nature of the  outflowing currents.
If indeed there are regions of antiparticle currents, we must make sure to modify the sign of $q$ appropriately in that region.
\section{The Lorentz Factor At Large Distances}
Since the Boyer-Lindquist coordinates are asymptotically flat, from the $t$ component of $W_I$ in eq.(\ref{reigionIW}), we see that the lorentz factor $\Gamma$ of the ejected mass at infinity becomes
$$\Gamma (\theta) = \frac{3l-1}{\sqrt{3l(3l-2)}}\;.$$
The above equation can be inverted to give
$$l = \frac{1}{3}\;\left[1 + \;\frac{\Gamma}{\sqrt{\Gamma^2-1}}\right]\;.$$
As $\Gamma \rightarrow 1$ the energy collimation factor $l \rightarrow \infty$, and when $l \rightarrow 2/3$ the Lorentz factor $\Gamma \rightarrow \infty$. In particular, for any finite $l$, the ejected particles are so energetic that it never comes to rest even infinitely far away from the black hole. Clearly, the collimation effects on $\Gamma$ stem from our freedom in choosing a judicious $l$. The strength of the jets at the poles are also compensated by the intensity of the emitted particles.

\section{ Symmetry Properties of the Force-Free Equations}

Consider a complete description of the fields and currents given by quantities $\rho$, $J$, $E$, $D$, $B$, and $H$. For each of these quantities will be define a dual object (the dual of a quantity $A$ will be indicated by $\tilde A$) such that all the dual objects collectively describe a force-free, stationary, axisymetric magnetosphere in Kerr geometry. The dual quantity will be very simply related to the original quantity, and yet, collectively, the physical content of the dual solutions will not be equivalent to the original one. The general features of this construction involve looking at the poloidal components of an object separately from the toroidal and the zeroeth component of the covariant formalism. This reason for this should be fairly clear: much like the background geometry, our assumptions require quantities to be time-independent (affecting the zeroeth component of a vector) and axisymetric (affecting the toroidal component of a vector).
\vskip0.2in
The dual charges and current are defined by

\begin{equation}
\tilde \rho =  \rho \;, \; \tilde J_T =  J_T \;, \; \tilde J_P = -J_P.
\end{equation}
Therefore if we define

\begin{equation}
\tilde D = D,
\end{equation}
we see that Gauss's theorem will be naturally satisfied since
\begin{equation}
\nabla \cdot \tilde D = \nabla \cdot D =\rho = \tilde \rho.
\end{equation}
For the other inhomogenous Maxwell's equation to hold, define
\begin{equation}
\tilde H_P =  H_P \;, \; \tilde H_T =  -H_T \;.
\label{newh}
\end{equation}
Then
\begin{equation}
(\nabla \times \tilde H)_T = (\nabla \times \tilde H_P)_T = (\nabla \times  H_P)_T = J_T = \tilde J_T,
\label{inhom1}
\end{equation}
and
\begin{equation}
(\nabla \times \tilde H)_P = (\nabla \times \tilde H_T)_P = (\nabla \times -H_T)_P = -J_P = \tilde J_P.
\label{inhom2}
\end{equation}
The second equality in the above equation holds only because the fields are axisymetric, for example
\begin{equation}
(\nabla \times \tilde H)^r =e^{r\theta\varphi} (\partial_\theta \tilde H_\varphi-\partial_\varphi \tilde H_\theta).
\end{equation}
Therefore, for time-independent solutions it follows from eqs.\ (\ref{inhom1}) and (\ref{inhom2}) that
\begin{equation}
-\partial_t \tilde D +  \nabla \times \tilde H = \tilde J.
\end{equation}
Now lets consider the homogenous Maxwell's equations. Having defined $\tilde H$ and $\tilde D$, we have no more freedom in picking $\tilde B$ and $\tilde E$. It is not difficult to see that
\begin{equation}
\tilde B_P =  B_P \;, \; \tilde B_T =  -B_T \;,
\label{newb}
\end{equation}
and
\begin{equation}
\tilde E =  E.
\end{equation}
Clearly, the curl of $\tilde E$ vanishes, and once again due to axis-symmetry, the divergence of $\tilde B$ is trivial as well. Therefore, we have shown that the new quantities satisfy Maxwell's equations. It is just a matter of simple calculation to show that the dual fields and currents are force-free. Therefore, there exists a 3-vector $\tilde\omega$ such that $\tilde E = -\tilde \omega \times \tilde B$. It turns out that $\tilde\omega = \omega$.

From eq.(\ref{newh}) we see that if $H_\varphi $ satisfies the Znajek regularity condition, $\tilde H_\varphi$ will not (unless $\tilde H_\varphi = - H_\varphi = 0$). Therefore, if we are using the dual solution to describe the external magnetosphere, we must separate the regions at $r = r_+ + \delta$ as explained in subsection \ref{outeventreg}.

\section{The  $\tilde \Omega_-$ Solution}
The dual solution to the $ \Omega_-$ solution presented earlier is given by
\begin{equation}
\tilde \Omega_- = \frac{1}{a \sin^2 \theta}
\end{equation}
\begin{equation}
\tilde E_\varphi = 0 = \tilde E_r \;,
\end{equation}
\begin{equation}
\tilde E_\theta = - \frac{2}{a^2}\; \Lambda \;\frac{\cos{\theta}}{\sin^5{\theta}}\;,
\end{equation}
and
\begin{equation}
\tilde B^\theta=0
\end{equation}
\begin{equation}
\tilde B^r=\alpha\;\tilde H^r = \frac{2}{a} \;\Lambda \;\frac{\cos{\theta}}{\sqrt{\gamma}\sin^3{\theta}}\;,
\end{equation}
and
\begin{equation}
\alpha\;\tilde B_\varphi = \tilde H_\varphi= - \frac{2}{a^2}\; \Lambda\; \frac{\cos\theta}{\sin^4\theta}\;.
\label{finalhphi}
\end{equation}
Finally,
\begin{equation}
\tilde I^{\nu} =- \frac{2}{a^2 \alpha \sqrt\gamma} \frac{d}{d \theta}[\Lambda \frac{\cos{\theta}}{\sin^4{\theta}}]\;l^{\nu},
\end{equation}
\vskip0.2in\noindent
where $l$ is the principle outgoing null geodesic of the Kerr geometry given by
\begin{equation}
l^{\nu} = \left(\frac{r^2+a^2}{\Delta}, 1, 0, \frac{a}{\Delta}\right).
\end{equation}
\vskip0.15in \noindent We will refer to the set of equations above as the $\tilde \Omega_-$ solution.
\subsection{Extraction Of Energy From The Electromagnetic Poynting Flux}
These solutions do allow for the extraction of energy  via the electromagnetic Poynting flux from the black hole.
From eq.(\ref{engextractformula}) we see that the rate of energy extraction

\begin{equation}
 \frac{d \tilde {\cal E}_{EM}}{dt} =-\int (-H_\varphi)  \Omega   B^r \sqrt {\gamma_{rr}} dA\ = - \frac{d {\cal E}_{EM}}{dt}\; >\;0\;.
\label{engext}
\end{equation}
\vskip0.2in\noindent
For the case of our $\tilde \Omega_-$ solution, the above expressions gives that

\begin{equation}
\frac{d {\tilde{\cal E}}_{EM}}{dt}=\frac{8\pi}{a^4} \int_{0}^{\pi} \frac{\Lambda^2\; \cos^2\theta}{\sin^9\theta} \;d\theta \geq 0 \;.
\end{equation}
\vskip0.2in\noindent
The collimation effects from the electromagnetic fields are given by

\begin{equation}
\frac{d^2 {\tilde{\cal E}}_{EM}}{dA\;dt}\;\approx \;\frac{4}{a^4}  \frac{\Lambda^2\; \cos^2\theta}{\sin^{10}\theta}\;\frac{1}{r^2 } \;.
\end{equation}

\subsection{Extraction Of Energy From The Electromagnetic Currents}

In region I, the dual  currents can be decomposed as follows:

$$\tilde I_I = \tilde I_{Ia} + \tilde I_{Ib} $$
where

$$\tilde I_{Ia} = \frac{2\;[3l -2]}{a^2\;\sin\theta \sqrt{\rho^2 \Delta}} \;\;\frac{d}{d \theta}\left[\Lambda \frac{\cos{\theta}}{\sin^4{\theta}}\right] \;V_I$$
and

$$\tilde I_{Ib} = -\frac{2\sqrt{3l(3l-2)}}{a^2\;\sin\theta \sqrt{\rho^2 \Delta}} \;\;\frac{d}{d \theta}\left[\Lambda \frac{\cos{\theta}}{\sin^4{\theta}}\right]\;W_I \;.$$
\vskip0.2in\noindent
Naturally, the extraction of matter energy stems from $\tilde I_{Ib}$. Analogous to eq. (\ref{matengexteqn}), here the matter extraction of energy $\tilde {\cal E}_M$ from the black hole is such that

$$\frac{d^2\tilde {\cal E}_M}{dt\;dA}  \approx  -\frac{m}{q}\;\frac{2}{a^2}\;\frac{d}{d \theta}\left[\Lambda \frac{\cos{\theta}}{\sin^4{\theta}}\right] \;\frac{1}{r^2\sin\theta}$$
and
$$\frac{d}{dt}\; \tilde {\cal E}_M= -\frac{4\;\pi}{a^2}\;\int \; \frac{m}{q}\;\frac{d}{d \theta}\left[\Lambda \frac{\cos{\theta}}{\sin^4{\theta}}\right]\; \; d\theta$$
\vskip0.2in\noindent
which is positive when $q(\theta)$ is correctly chosen so that

\begin{equation}
-\frac{1}{q}\;\frac{d}{d \theta}\left[\Lambda \frac{\cos{\theta}}{\sin^4{\theta}}\right] \geq 0\;.\label{tilchgsgneqn}
\end{equation}
\section{A Particular Choice Of Lambda}
Since we want energy extraction via both the electromagnetic fluxes and the matter currents, through out this section, we will focus on the $\tilde \Omega_-$ solution. Consider the simplest case where
\begin{equation}\Lambda = \Lambda_0 \; \sin^5 \theta \;.\label{specificlam}\end{equation}
The factor of $\sin^5 \theta $ is necessary to make the fields well defined on the poles.
From eq. (\ref{tilchgsgneqn}) we have that
$$q \equiv q_- < 0\;\;\;{\rm when}\;\;\; 0 \leq \theta < \pi/4\;, \;\;\;{\rm and}\;\;\; \frac{3\;\pi}{4} < \theta \leq \pi \;.$$ and
$$q \equiv q_+ > 0\;\;\;{\rm when}\;\;\; \pi/4 < \theta < 3\;\pi / 4 $$
 when $\Lambda_0 $ is positive, which we will now require. Let $m_-$ and $m_+$ be the corresponding mass of the particle species. Then

$$\frac{d^2\tilde {\cal E}_M}{dt\;dA}  \approx  - \frac{m}{q}\;\frac{2 \;\Lambda_0 }{a^2} \;\frac{\cos 2\theta}{r^2\;\sin \theta}\;.$$
\vskip0.2in\noindent
Clearly, this describes a jet like solution where the rate
of energy extraction is maximized at the poles. There is a second local maxima along the equatorial plane suggesting the existence of a strong accretion disk. The above expression is integrated to give
$$\frac{d}{dt}\; \tilde {\cal E}_M= \frac{4\;\pi\;\Lambda_0}{a^2}\left[ \frac{m_+}{q_+} - \frac{m_-}{q_-}\right]\;.$$
We are not assuming that the species of charged particles in the two regions (near the poles and near the equatorial planes) are particle antiparticle pairs, although nothing precludes it in our formalism. They in could in particular be currents of electrons and protons (so that the black holes remains neutral during the process of energy extraction). The electromagnetic Poynting flux however gives that
\begin{equation}
\frac{d^2 \tilde {\cal E}_{EM}}{dA\;dt}\;\approx \;\frac{4}{a^4} \;\frac{\Lambda_0 ^2\;\cos^2 \theta}{r^2 } \;.
\end{equation}
Here, just as in the case of the matter currents, the rate of electromagnetic energy extraction is a maximum along the polar axis. Interestingly, at the equatorial plane, the electromagnetic extraction rate is trivial suggesting a secondary mechanism for the observed glow around the accretion disk.  The total extraction rate is given by
\begin{equation}
\frac{d \tilde {\cal E}_{EM}}{dt}=\frac{16\;\pi\;\Lambda_0 ^2}{3\;a^4} \;.
\end{equation}

\section{Magnetic Monopoles}
Since $\tilde B^r = B^r$, as long as $\Lambda$ is the same in regions I and II, we will not introduce any magnetic monopoles as a result of using the dual solution in the external region I ($r > r_+ +\; \delta$). However, it is important to pick $\Lambda$ carefully if we are to exclude magnetic monopoles in every closed region of spacetime. Our choice of $\Lambda$ given by eq. (\ref{specificlam}) is one such choice. Indeed

$$\int_{r={\rm const}} g(B, n)\; dA = \frac{4 \;\pi}{a}\int_0 ^\pi \Lambda \frac{\cos\theta}{\sin^3 \theta}\;d\theta$$$$ = \frac{4 \;\pi\;\Lambda_0}{a}\int_0 ^\pi \cos\theta \;\sin^2 \theta\;d\theta = 0\;.$$


\section{Conclusion}
There are two important clarifications that we must make before we conclude this article. In \cite{Ref11}, we had claimed that it is impossible to extract energy when $\Omega = \Omega_-$. The reference in this case is to the energy extracted via the electromagnetic flux alone; and it still holds true here (as it must). In this paper, as we have seen, it is possible to extract energy via matter currents when $\Omega = \Omega_-$. The extraction of energy via electromagnetic flux also does occur when $\Omega = \tilde \Omega_-$. The price to pay in this case is in the discontinuity of $H_\varphi$. This in turn produces a delta function current at the membrane joining the two regions. The analysis of this current on the membrane requires further study.

Outside of a few mild constraints, the functions $l$ and $\Lambda$ are arbitrary functions of $\theta$. It is not clear whether astrophysical black holes permit a wide variety of magnetospheres, or if there is some other mechanism restricting the large degrees of freedom the decomposed currents and the fields have. Nonetheless, we have constructed a specific, exact solution to the Blanford-Znajek mechanism that extracts energy from the black hole.

\section*{Acknowledgments}
The first author would like to thank Troy University for their continued support of our research in black hole astrophysics, while the office of Naval research funds and supports the second author.

\appendix

\section{Kerr Geometry Essentials}
\label{KGE}
For completeness, we define the various Kerr coordinates used. For asymptotic analysis, the Boyer-Lindquist coordinates are preferred, while the horizon and the interior region ($r \leq r_+$) is analyzed using the usual Kerr-Schild coordinate system.
\subsection{ Boyer-Lindquist Coordinates}
In the Boyer-Lindquist
coordinates $\{t,r,\theta,\varphi\}$ of the Kerr geometry, the metric takes the form
$$
ds^2=( \beta^2 - \alpha^2 )\;dt^2 \;+  \;2 \;\beta_\varphi \;d\varphi
\;dt$$$$
+\gamma_{rr}\; dr^2 + \;\gamma_{\theta \theta}\; d\theta^2 +
\;\gamma_{\varphi\varphi}\;d\varphi^2 \;,$$
where the metric coefficients are given by
$$\beta^2-\alpha^2 \;= \;g_{tt} \;=\; -1 + \frac{2Mr}{\rho^2}\;,$$$$\beta_\varphi \; \equiv g_{t \varphi}\; = \;\frac{-2Mr a
\sin^2\theta}{\rho^2}\;,\;\;\;\gamma_{rr} =
\frac{\rho^2}{\Delta}\;,$$
$$
\gamma_{\theta \theta} = \rho^2, \;\; {\rm and}\;\; \gamma_{\varphi \varphi} = \frac{\Sigma^2 \sin^2\theta}{\rho^2}\;.
$$
Here,
$$\rho^2 = r^2 + a^2
\cos^2\theta\;,\;\;\;\Delta = r^2 -2 M r + a^2$$
and
$$
\Sigma^2 = (r^2 + a^2)^2 -\Delta \; a^2 \sin^2\theta\;.
$$
 Additionally
$$
\alpha^2 = \frac{\rho^2 \Delta}{\Sigma^2}, \;\;\; \beta^2 =
\frac{\beta_\varphi^2}{\gamma_{\varphi \varphi}}$$
and
$$\sqrt{-g}=\alpha\; \sqrt{\gamma} = \rho^2 \sin\theta\;.$$
The parameters $M$ and $a$ are the mass and angular momentum per unit
mass respectively of the Kerr black hole. The horizons $H_\pm$  are located at   $ r_\pm = M \pm \sqrt{M^2 - a^2} $.
\subsection{ Kerr-Schild Coordinates}
Kerr-Schild coordinates are given by the transformation

\begin{equation}
\left[\begin{array}{c}
d\bar t\\
d \bar r\\
d \bar \theta\\
d \bar \varphi\\
\end{array}\right]
=\left[\begin{array}{cccc}
1 & G& 0& 0\\
0& 1& 0& 0\\
0 & 0 & 1 & 0\\
0& H& 0& 1\\
\end{array}\right]
\left[\begin{array}{c}
d t\\
d r\\
d \theta\\
d \varphi\\
\end{array}\right],
\label{transformup}
\end{equation}
where
\begin{equation}
G = \frac{r^2+a^2}{\Delta}\;\;\;\;\; {\rm and}\;\;\;\;\; H = \frac{a}{\Delta}\;.
\end{equation}
In this frame, the metric becomes
\begin{equation}
 g_{ \mu  \nu} = \left[\begin{array}{cccc}
z-1& 1& 0& -za\sin^2\theta\\
1& 0& 0& -a\sin^2\theta\\
0 & 0& \rho^2 & 0\\
-za\sin^2\theta & -a\sin^2\theta& 0& \Sigma^2 \sin^2\theta/\rho^2\\
\end{array}\right] \;,
\label{kerrschildmetric}
\end{equation}
where $z = 2Mr/\rho^2$.
Components of vectors transform as
$$\bar X^\mu = A^\mu ~_\nu\;\;X^\nu$$
and 1-forms transform as
$$\bar X_\mu = \left [(A^{-1})^T \right ]_\mu ~^\nu\;X_\nu$$
where the matrix $A$ is given in eq.(\ref{transformup}).

\vskip0.2in
We pick our time orientation for the Kerr geometry such that the null vector field $-\partial_{\bar r}$ is future pointing everywhere.

\section{ Equations Of Electrodynamics In Stationary Spacetimes}
\label{EESS}

We only state the relevant equations of electrodynamics of stationary spacetimes. For a detail development, see \cite{DMbook}. Maxwell's equations can be written as
\begin{equation}
\nabla_\beta \star \;F^{\alpha \beta} = 0 \;, \;{\rm and} \;
\nabla_\beta  F^{\alpha \beta} = I^\alpha\;.
\label{maxeq}
\end{equation}
Here $F^{\alpha \beta}$ is the Maxwell stress tensor,  $I^\alpha$ is the four vector of the electric current  and $\nabla$ is the covariant derivative of the geometry. $\star \;F$ is
the two form defined by
\begin{equation}
\star \;F^{\alpha \beta} \equiv \frac{1}{2}\epsilon^{\alpha \beta \mu \nu} F_{\mu \nu}\;.
\end{equation}
Here, $\epsilon_{\alpha \beta \mu \nu}$ is the completely antisymmetric Levi-Civita tensor density of spacetime such that $
\epsilon_{0123}= \sqrt{-g}=  \alpha \sqrt{\hat \gamma}
$ ($\alpha$ and $\gamma$ along with the other relevant Kerr quantities are defined in section \ref{KGE}. In the 3+1 formalism, where $\partial_0$ is the asymptotically stationary timelike killing vector field, $ E$ and $ B$ are defined so that
\begin{equation}
 F_{ \mu  \nu} =
\left[\begin{array}{cccc}
0&  -E_1&  -E_2 & -E_3\\
  E_1 & 0 & \sqrt{\gamma}\; B^3 & - \sqrt{\gamma}\; B^2\\
E_2& - \sqrt{\gamma} \;B^3 & 0 & \sqrt{\gamma} \;B^1\\
E_3 & \sqrt{\gamma}\;B^2 & - \sqrt{\gamma} \;B^1 & 0\\
\end{array}\right] \;.
\label{fdown}
\end{equation}
We also define dual vectors $ D$ and $ H$ by
\begin{equation}
 * F_{ \mu  \nu} =
\left[\begin{array}{cccc}
0&  H_1&  H_2 & H_3\\
-  H_1 & 0 & \sqrt{\hat\gamma}\; D^3 & - \sqrt{\hat\gamma}\; D^2\\
- H_2& - \sqrt{\hat\gamma}\; D^3 & 0 & \sqrt{\hat\gamma} \;D^1\\
- H_3 & \sqrt{\hat\gamma} \; D^2 & - \sqrt{\hat\gamma} \; D^1 & 0\\
\end{array}\right] \;.
\label{starfdown}
\end{equation}
Naturally, $F$ and $\star \;F$ are not independent. They are related by
\begin{equation}
\alpha D = E- \beta \times B
\label{contitutive1}
\end{equation}
and
\begin{equation}
H = \alpha B - \beta \times D\;.
\label{contitutive2}
\end{equation}
Here,
\begin{equation}
(A \times B)^i \equiv \; \epsilon^{ijk}  \; A_j\; B_k\;,
\end{equation}
where $\epsilon^{ijk}$ is the Levi-civita tensor of our absolute space defined $x^0 = {\rm constant}$. Also, $\beta$ is the shift dual vector given by $\beta = \beta_\varphi \;d\varphi$. Naturally, the spatial coordiantes are given by $(x^1, x^2, x^3)$, and three vectors $ E, B, D, H$ live in this absolute space. Now, Maxwell's equations can be re-written as
\begin{equation}
\tilde \nabla \cdot B = 0\;,
\label{divb}
\end{equation}
\begin{equation}
\partial_t B + \tilde \nabla \times E = 0\;,
\label{faraday}
\end{equation}
\begin{equation}
\tilde \nabla \cdot D = \rho_c\;,
\label{maxcharge}
\end{equation}
and
\begin{equation}
-\partial_t D + \tilde \nabla \times H = J\;,
\label{maxcurrent}
\end{equation}
where $\rho_c = \alpha I^t$ and $J^k = \alpha I^k$. Here $\rho_c$ is the charge density and $J$ is the electric 3-current. $\tilde \nabla$ is the covariant of the 3 space with the induced metric.
The force-free condition that we will enforce is
\begin{equation}
F_{\nu \alpha}\; I^\alpha =0\; .
\label{divtemforce}
\end{equation}
This condition takes the form
\begin{equation}
E \cdot J = 0
\label{fofree1}
\end{equation}
and
\begin{equation}
\rho_c E + J \times B = 0.
\label{fofree2}
\end{equation}
For the case of a stationary,  axis-symmetric,  force-free magnetosphere, it is easy to show that there exists $\omega = \Omega\;\partial_{\varphi}$ such that
$$E = - \omega \times B\;.$$
Additionally,  \cite{Ref5} showed that
\begin{equation}
H_\varphi \left |_{r_+} = \frac{\sin^2\theta}{\alpha}\; B^r\; (2Mr\; \Omega -a) \right |_{r_+}
\label{znaregcond}
\end{equation}
is the required condition in the Boyer-Lindquist coordinates that the otherwise bounded fields must satisfy so that they continue to be well defined in the Kerr-Schild coordinates at the event horizon. Eq.(\ref{znaregcond}) is referred to as the Znajek regularity condition.

\label{lastpage}

\end{document}